\newcommand{\q}[1]{``#1''}

\documentclass[conference]{IEEEtran}
\IEEEoverridecommandlockouts
\usepackage{cite}
\usepackage{amsmath,amssymb,amsfonts}
\usepackage{algorithmic}
\usepackage{graphicx}
\usepackage{textcomp}
\usepackage{xcolor}
\begin{document}

\title{Classification of Electroencephalograms during Mathematical Calculations Using Deep Learning}

\makeatletter
\newcommand{\linebreakand}{%
  \end{@IEEEauthorhalign}
  \hfill\mbox{}\par
  \mbox{}\hfill\begin{@IEEEauthorhalign}
}
\makeatother
\makeatletter
\newcommand\thefontsize{\expandafter\string\the\font}

\author{\IEEEauthorblockN{Umang Goenka\IEEEauthorrefmark{1}, Param Patil\IEEEauthorrefmark{2}, Kush Gosalia\IEEEauthorrefmark{3}, Aaryan  Jagetia\IEEEauthorrefmark{4}}
\IEEEauthorblockA{\IEEEauthorrefmark{1}\IEEEauthorrefmark{4}Department of  Information Technology,
Indian Institute Of Information Technology, Lucknow, India\\
\IEEEauthorrefmark{2}Department of  Computer Engineering,
Sardar Patel Institute Of Technology, Mumbai, India\\
\IEEEauthorrefmark{3}Department of Mechanical Engineering,
Indian Institute Of Technology Patna, India\\
\IEEEauthorrefmark{1}lit2019033@iiitl.ac.in,
\IEEEauthorrefmark{2}param.patil@spit.ac.in,
\IEEEauthorrefmark{3}kush\_1901me37@iitp.ac.in,
\IEEEauthorrefmark{4}lit2019045@iiitl.ac.in}}

\maketitle

\begin{abstract}
Classifying Electroencephalogram(EEG) signals helps in understanding Brain-Computer Interface (BCI). EEG signals are vital in studying how the human mind functions. In this paper, we have used an Arithmetic Calculation dataset consisting of Before Calculation Signals (BCS) and During Calculation Signals (DCS). The dataset consisted of 36 participants. In order to understand the functioning of neurons in the brain, we classified BCS vs DCS. For this classification, we extracted various features such as Mutual Information (MI), Phase Locking Value (PLV), and Entropy namely Permutation entropy, Spectral entropy, Singular value decomposition entropy, Approximate entropy, Sample entropy. The classification of these features was done using  RNN based classifiers such as LSTM, BLSTM, ConvLSTM, and CNN-LSTM. The model achieved an accuracy of 99.72\% when entropy was used as a feature and ConvLSTM as a classifier.

\end{abstract}


\section{INTRODUCTION}
The mechanism of the human brain in today's date is still unexplored and not known to the researchers. Every action we take has its own set of signals depending on which the human brain reacts.
The understanding of these signals is something that is fundamental in the process to decipher how the human brain functions [1].

Signals are processed with an equipment or a conventional methodology called EEG, which will help us to visualize those signals. EEG [2] is simply a method to track the electrical signals of the human brain. They consist of electrodes that are shaped in the form of a small metal disc with thin wires. Those electrodes basically record the signals passing through them and reflect it in the form of graphs. Generally a set of 21 electrodes are used for standard purposes as mentioned in Fig. 1.
\begin{figure}[htbp]
\centerline{\includegraphics[width=5cm, height=4cm]{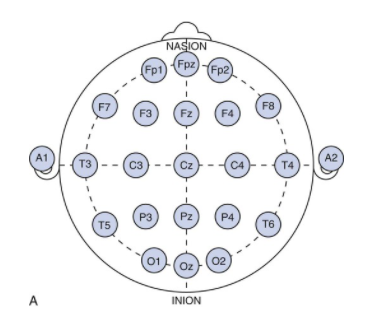}}
\caption{Position of 21 channels for recording EEG signal}
\label{fig}
\end{figure}
The rate at which the information is being processed by a person could be directly proportional to the fluctuations in the signal graphs represented by the EEG devices. But this analogy may vary depending on the stress a person is put into. There are many ways a human could be put under pressure, but the method that we chose was to ask Arithmetic questions to a person whose difficulty level may gradually increase and accordingly the signals or in layman terms the information passed from one neuron to another could be shown on a graph.

Keeping a track of the human’s cognitive workload is the first step but to analyze it so that we can later on classify them is the next step. Earlier a simple questionnaire based methodology was executed, where the users frustration or pressure were simply evaluated into a metris to establish a connection with the performance of a task. But with the evolution of technology and introduction of Machine Learning/Deep Learning in almost every sector, the efficiency of evaluating these cognitive workloads has also increased. 

With the help of state of the art Deep Learning algorithms like recurrent neural network(RNN) and libraries like keras, using various long short-term memory(LSTM), bi-directional LSTM(BLSTM), Convolutional LSTM(ConvLSTM) based neural network architectures, we try to classify the signals into two categories i.e if the user is performing a mental calculation [3] or not. Unlike ordinary data processing, where raw data can also help us to get good accuracy, EEG requires certain feature extraction methods and so we tried to construct MI, PLV matrix. In recent times, entropy has also helped in improving the results and therefore we did feature extraction by dividing these time series dataset into various batches of 2sec, 3sec, 4sec before we fetch them into our DL model.


\section{background and related work}
In this section, we review, summarize and analyze work related to EEG signal classification using Machine Learning techniques.

 Abhishek Varshney et al. [4] extracted features such as the approximation entropy, sample entropy, permutation entropy, dispersion entropy, and slope entropy. This paper used LSTM, BLSTM, and Gated Recurrent Unit(GRU) for classification of the cognitive workload tasks.

 Qiang Wang et al. [5]  proposed a methodology to classify real-time EEG signals using multifractal analysis for arithmetic task recognition. This paper extracted features such as power spectrum density, autoregressive model. The fractal dimension was defined as the combination of all the features and classified using Support Vector Machine (SVM).

Binish Fatimah et al. [6] used features like energy entropy, mean and L2 norms were extracted by the rhythms filtered using the EEG signals. This paper used SVM, decision tree and quadratic discriminant analysis for classification along with entropy for feature extraction.

Biswarup Ganguly et al. [7] presented an EEG based mental arithmetic task classification to study brain computer interfacing (BCI). EEG signals from 36 subjects were recorded and eight features were extracted per electrode. These features were fed to a stacked long-short term memory (LSTM) architecture for enhancing and building the brain computer interfacing model.

B. Rebsamen et al. [8] proposed a start towards building a real-time system which measures cognitive workload from EEG. Data analysis and classification was done using a quadratic discriminant analysis classifier in PyMVPA. 

Kenji Katahira et al. [9] presented a methodology to establish a dependable measurement of the flow state by using EEG in the presence of an experimental flow state. Mental arithmetic tasks on Boredom, Flow and Overload were given to perform to 16 participants - 10 men and 6 women. The EEG data was analyzed using EEGLAB and FieldTrip and correlations were made between subjective flow and EEG data.

Binish Fatimah et al. [10] proposed a mental arithmetic task detection algorithm which uses Fourier decomposition to understand the brain response from a single lead EEG signal. Arithmetic tasks (serial subtraction of two numbers) were conducted for women aged 16 to 21 and men aged 17 to 26. The decomposed signals were used to extract features like energy, entropy and variance and classified using SVM. 


Soo-In Choi et al. [11] proposed the use of Ear-EEG to develop endogenous BCI systems which use self-modulated brain signals. EEG data was collected from seven participants aged between 21 and 31 who performed mental arithmetic (MA) and baseline (BL) tasks. EEG features were extracted by applying the common spatial pattern (CSP) to the data.

Rifai Chai et al. [12] presented EEG based BCI which uses the prefrontal cortex non-hair area for classification. Five subjects including 3 males and 2 females aged between 25 and 35 years performed mental tasks including baseline, arithmetic calculation, finger tapping, ringtone and words association. Features were extracted using the Hilbert Huang Transform (HHT) energy method and classified using an artificial neural network (ANN) with genetic algorithm (GA) optimization. 

Few papers mentioned above have used the similar feature extraction technique to get the accuracy, in our paper we have also considered few more entropy parameters like permutation entropy, SVD(single value decompostion entropy). The parameters which were used in this paper have helped us to get better results with different combinations of the input shape given to the model. Our paper also focuses on how the input shape given to the model with data that has gone through various feature extraction method can provide better accuracy on some methods and a average accuracy on others. The methods applied in this paper are extension to the proposed methods in the above papers with new entropy and model architecture.


    


\section{Methodology}

This section describes the feature extraction process and structure of the classifiers. The model then preprocesses the signals, extracts the features, and classifies them into BCS and DCS as referred in Fig. 2.
\begin{figure}[htbp]
\centerline{\includegraphics[width=6cm, height=11cm]{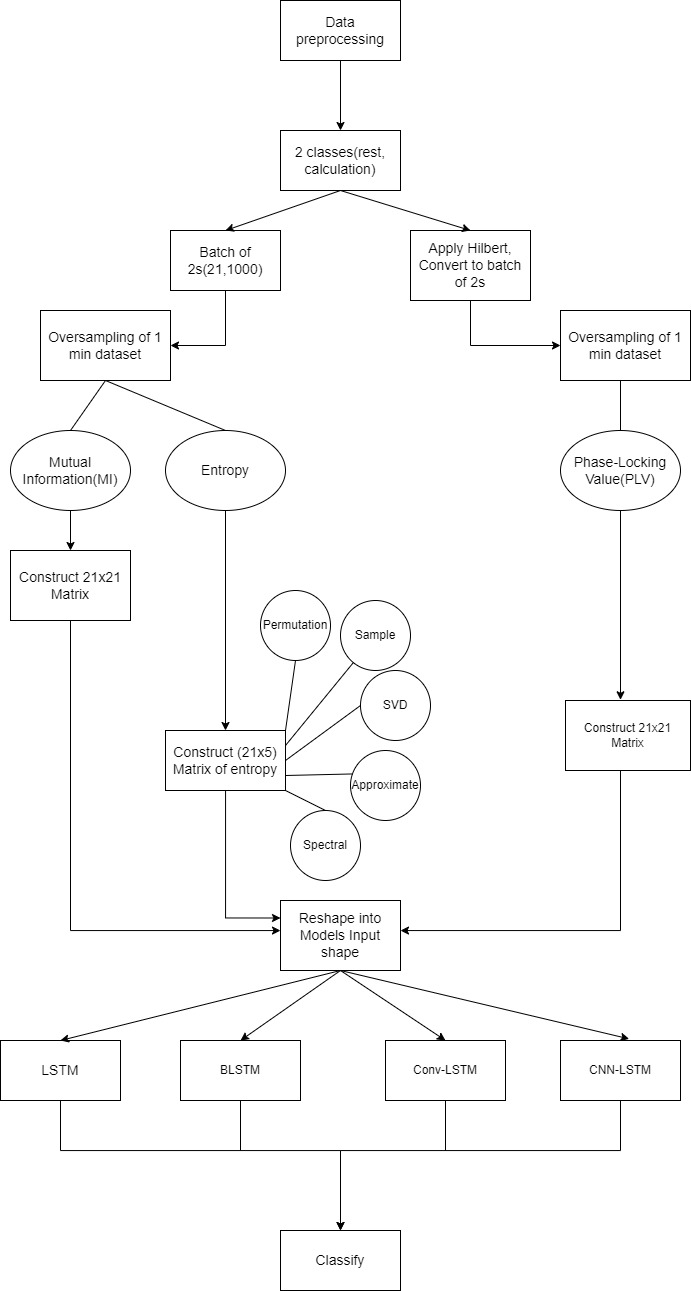}}
\caption{System Architecture}
\label{fig}
\end{figure}

\subsection{Dataset}\label{AA}

The dataset [13] consists of 36 participants whose signals were recorded. The recording of 180 seconds of BCS and 60 seconds of DCS. Each recording has 21 channels as mentioned in Fig. 3. The electrodes are placed such that they can capture as many signals as possible. Each subject is told to do an n-subtraction task. Initially each subject is given two variables x and y. The subject has to subtract y from x as many times as possible and remember the result. The subject does this task for 1 minute continuously. At the end of the task it is checked whether the subtraction operation was done correctly or not by calculating the remainder.


\begin{figure}[htbp]
\centerline{\includegraphics[width=6cm, height=4cm]{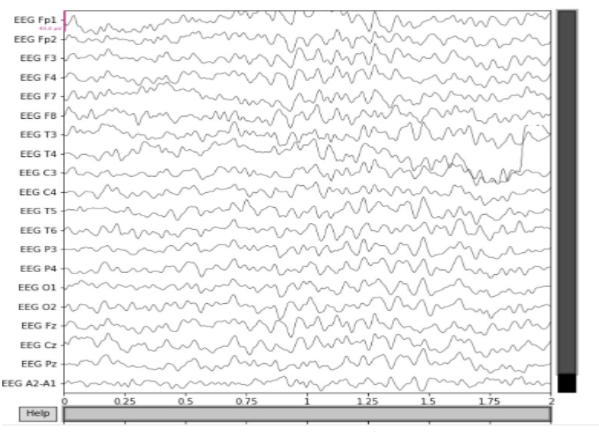}}
\caption{EEG signal for a 2-second frame for DCS}
\label{fig}
\end{figure}

\subsection{Feature Extraction}

1) Phase Locking Value:

PLV[14] helps in investigating task-induced changes in long-range synchronization of neural activity from EEG data. The two signals which are going to be input must be first transformed using Hilbert Process. Here
\(\Delta\psi_{k,l}(t)\) is the phase difference of the two signals k and l. The formula is given in Equation below.
\[1: PLV(k,l) = \frac{1}{n_t} |\sum_{t=1}^{n_t} e^{j\Delta\psi_{k,l}(t)}|\]

2) Mutual Information:


MI[15] is used to relate the interdependence of two signals a and b. In MI we give two signals as input and calculate the joint probability function P(a,b). MI is calculated in Equation below.
\[2:MI_{ab} =\sum_{a\epsilon A,b\epsilon B} P(a,b)\log \frac{P(a,b)} {P(a)P(b)}\]

3) Entropy:

Here we calculated 5 different types of entropy [16]:
Approximate Entropy: Approximate Entropy[17] measures correlation in the sense that low Approximate Entropy values reflect that the system is very persistent with apparent patterns that repeat themselves throughout of the series, while high values mean independence between the data, a low number of repeated patterns and randomness.
\[3: \phi^m(r) = \frac{1}{N-m+1} \sum_{i=1}^{N-m+1} \log C_{i^m}(r)\]

\[4: ApEn(m,r)= \lim_{N\to\infty} [\phi^m(r)-\phi^{m+1}(r)]\] 
Sample Entropy: Sample Entropy[18] is a measure of complexity. It is slightly different from approximate entropy as it does not include self similar patterns. Here 

\[\mbox{A = number of template vector pairs having } \]
\[d[X_{m+1}(i),X_{m+1}(j)]<r\]
\[\mbox{B = number of template vector pairs having }\] 
\[d[X_{m}(i),X_{m}(j)]<r\]
\[5: SampEn(m,r,N) = - \log (\frac{\sum_{i=1}^{N-m} A_i}{\sum_{i=1}^{N-m} B_i})=-\log (\frac{A}{B})\]
Spectral Entropy: Spectral Entropy[19] is a normalized form of Shannon entropy, which uses the power spectrum amplitude components of the time series for entropy evaluation 
\[6: x_i = \frac{X_i}{\sum_{i=1}^N X_i} \mbox{ for i = 1 to N}\]
\[7: H(x)=-\sum_{x\epsilon X} x_i . \log_2 x_i\]
Permutation Entropy: Permutation Entropy[20] is a robust time series tool which provides a quantification measure of the complexity of a dynamic system by capturing the order relations between values of a time series and extracting a probability distribution of the ordinal patterns 
\[8: PE_{D,norm}=-\frac{1}{\log_2 D!}\sum_{i=0}^{D!}p_i\log_2 p_i\]
Singular Value Decomposition Entropy(SVD): SVD entropy [21] is an indicator of how many vectors are needed for an adequate explanation of the data set.  It measures feature in the sense that the higher the entropy of the set of SVD weights, the more orthogonal vectors are required to adequately explain it.
\[9: H_{SVD}=-\sum_{i=1}^M \bar{\sigma_i} \log_2 \bar{\sigma_i} \]
\[\mbox{where M is the number of singular values and } \sigma_1,...,\sigma_M\]
\[\mbox{are normalized singular values by function } \bar{\sigma_i}=\frac{\sigma_i}{\sum_{j=1}^M \sigma_j}\]
\newline



PLV is a standard feature extraction method which have been in use for a long time. We decided to perform our analysis using Mutual Information because it has been generally used to study cause-effect relationship in EEG. We used entropy which is considered to be more robust in this paper and wanted to compare the results with PLV and MI to get an understanding of the efectiveness of entropy over other commonly used methods.

\subsection{Data Preprocessing}

Here the data which we got from the dataset consisted of numbers and for each person a file with 3 mins data while resting and 1 min data while performing calculation. Each person's 3 mins data roughly translates to 90000 rows and 21 columns(each channel/electrode is considered as a single column) and 1 min data to 30000 rows and 21 columns with a frequency of 500Hz.

We chose mne for reading the raw data, we read the raw data for and reshaped them for MI and Entropy, and applied Hilbert process before reading the raw data for PLV.

We try to break each files data of every person into batches of 2s, considering the frequency as 500Hz, the total time points in a single batch would be 2*500 = 1000. Therefore the shape of each batch was (21, 1000). Some of the files in 3 mins data consisted of 40000 rows or 1,00,000 rows, but after breaking them into batches of 1000, most of the files had a shape to (91,21,1000) and after merging all the persons data into a single array the final shape was found out to be (36,91,21,1000).
On the other hand, files with 1 min data consisted and thus after breaking them into batches of 1000, the shape of each file translated to (31, 21, 1000) and after merging all the persons data the final shape was (36, 31, 21, 1000) and after oversampling it was (36*31*3, 21, 1000). 

At the end we did a 80-20 split of the above shape to obtain training and testing dataset. \newline

1) Feature vs Model Input Table:
\begin{table}[htbp]
    \caption{Dimension of features for different models}
    \begin{center}
    \begin{tabular}{ |p{1cm}|p{1.3cm}|p{1.3cm}|p{1.6cm}|p{1.3cm}|p{1.3cm}| }
     \hline
     & LSTM & BLSTM & ConvLSTM & CNN-LSTM \\
     \hline
     MI & (1,441) & (1,441) & (1, 21, 21, 1) & (21,21,1)\\
     \hline
     PLV & (1,441) & (1,441) & (1, 21, 21, 1) & (21,21,1)\\
     \hline
     Entropy & (1,105) & (1,105) & (1, 21, 5, 1)/ (1, 1, 105, 1) & (21,5,1)\\
     \hline
    \end{tabular}
    \end{center}
    \end{table}
\\
i) Mutual Information: Here we try to construct a 21x 21 Matrix with each column giving its own relationship with every other column. For a batch of 1000 rows, we use sklearns’ Mutual\_info\_score to generate a 21x21 matrix for a single batch of shape (21,1000).Later on while fetching into the model we either reshape it into (1,441) or (21,21) depending on the network’s feature property.\newline
ii) PLV:  Before constructing the PLV matrix the data has to be gone through a hilbert process. After that for every batch size of (21, 1000) we construct a 21x21 matrix by the following method:
 
Where theta 1, theta 2 are basically a single column/channels data in a batch with shape( 1000,). After that depending upon the network’s feature property we set input as (1, 441) or (21, 21).
\newline
iii) Entropy:  We considered 5 types of entropy as mentioned above in feature extraction and try to construct a (21, 5) matrix for a (21, 1000) batch size. The process that we go through is as follows:

We used the entropy library in python for the same which already had these built-in functions. The input given to the function is a single column data (1000,) and finally we get an array of size 5 which has an entropy data of various entropies mentioned earlier, thereby doing this with every column would in turn return us a shape of (21, 5). Ultimately, depending on the network’s feature properties, we either give (21, 5) as as input or (1, 21*5) as an input. 

\subsection{Neural Network:}
There were three different types of LSTM models that were used for classification purposes:
\begin{enumerate}
    \item 2 layered LSTM with 128 nodes and Dense layer of 32
    \item 2 layered LSTM with 64, 32 nodes and Dense layer of 32, 16
    \item 4 layered LSTM with 64, 32, 16, 16 nodes and Dense layer of 64, 16.
\end{enumerate}\par

Each of these models[22] were given an input feature of (1, 441), where 1 is the time\_steps and 441 = 21*21 is the total number of features obtained by flattening the PLV matrix. The best result was obtained from the second type of model in case of LSTM.
In case of BLSTM the model architecture was 2 layered BLSTM with 64, 32 nodes and Dense layer of 64, 16. The input feature in case of BLSTM was also the same i.e (1,441). For ConvLSTM the model architecture consisted of 2 2d ConvLstm layers with kernel size 3 and filter 64.First layer was followed by a BatchNorm before entering the second layer. The input shape given was (1, 21, 21, 1), where the first 1 represents the time\_steps , 21, 21 is a 2D matrix feature and the last 1 depicts the number of channels which is 1.
In the case of CNN-LSTM the input shape given was (21, 21, 1) for PLV, MI and (21, 5, 1) for Entropy features. The model first consists of a 2d Conv layer with kernel size 3 and filter 64, which were later on time distributed to ensure it follows the RNN structure followed by a Maxpooling of size 1. This pooling layer was followed by another 2D time distributed conv layer with filter 64 and kernel 3. After the dropout and flattening of the CNN layers, we add 2 LSTM layers of 32, 16 nodes in order. The model finally has a Dense layer of 100 which is followed by a softmax layer for classification.
One important thing to note is that at every layer except for the last for all the models the activation was relu. The optimizer was Adam with a learning rate of 10\^-3 and decay of 10\^-5.
As we focused on the classification of the signals, therefore the loss was also chosen as categorical cross entropy.

\subsection{Layer Diagram:}
The model architecture of all the 4 Neural Networks i.e. LSTM, BLSTM, ConvLSTM and CNN-LSTM are mentioned in Fig. 4. 
\begin{figure}[htbp]
\centerline{\includegraphics[width=8cm, height=10cm]{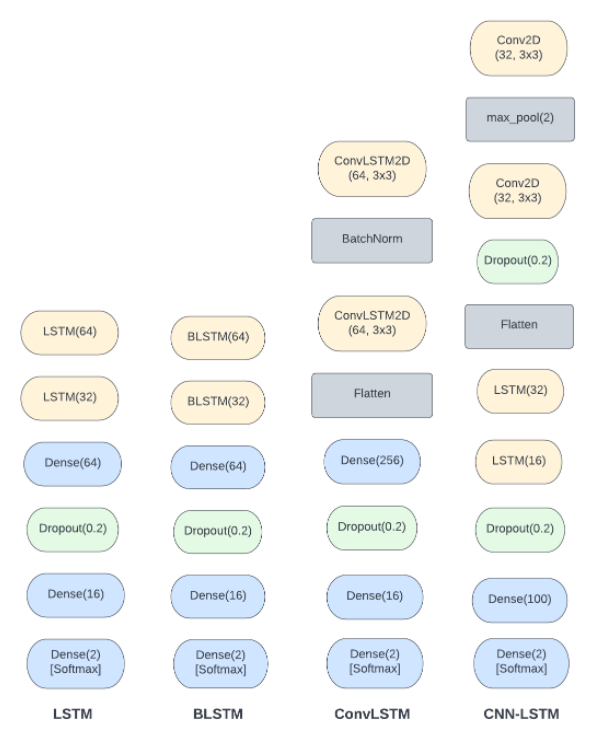}}
\caption{Layer Diagram}
\label{fig}
\end{figure}
\section{OBSERVATIONS AND RESULTS}
This section describes the various algorithms and architectures we experimented with.

\begin{enumerate}

    \item Phase Locking Value:
    \begin{table}[htbp]
    \caption{Accuracy of PLV}
    \begin{center}
    \begin{tabular}{ |p{2cm}|p{2cm}|p{2cm}|p{2cm}|  }
     \hline
     Model & Train Accuracy & Test Accuracy \\
     \hline
     LSTM  &  93.19\% &  77.33\%\\
     \hline
     BLSTM &  82.44\% & 80.73\%\\
     \hline
     Conv-LSTM  &  93.02\% &  77.33\%\\
     \hline
     CNN-LSTM  & 92.21\%  & 76.97\% \\
     \hline
    \end{tabular}
    \end{center}
    \end{table}
    
    \begin{figure}[htbp]
    \centerline{\includegraphics[width=6cm, height=4cm]{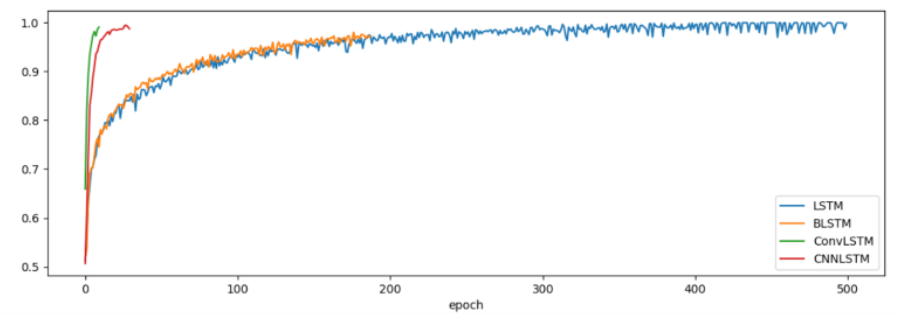}}
    \caption{Training graph of PLV}
    \label{fig}
    \end{figure}
    PLV showed a very low accuracy as mentioned in Fig. 5 on arithmetic calculation dataset. The maximum accuracy of 95.60\% was achieved using BLSTM. Also, while using PLV there was a lot of overfitting as there was nearly a 20\% gap between test and train accuracy as referred in table. 2. A lot of changes were made to the model architecture but the performance didn’t improve drastically. \newline
    \item Mutual Information:
    \begin{table}[htbp]
    \caption{Accuracy of MI}
    \begin{center}
    \begin{tabular}{ |p{2cm}|p{2cm}|p{2cm}|p{2cm}|  }
     \hline
     Model & Train Accuracy & Test Accuracy \\
     \hline
     LSTM  & 98.97\% & 94.11\%\\
     \hline
     BLSTM & 99.70\% & 97.45\%\\
     \hline
     Conv-LSTM & 99.59\% & 95.69\%\\
     \hline
     CNN-LSTM  & 99.69\%  & 97.01\%\\
     \hline
    \end{tabular}
    \end{center}
    \end{table}
    
    MI showed a lot better accuracy than PLV and also there is very less variation in test and train accuracy as mentioned in Table. 3. MI worked the best with BLSTM and gave the highest test accuracy of 97.45\%. The best training accuracy was 99.70\% as mentioned in Fig. 6. \newline
    \begin{figure}[htbp]
    \centerline{\includegraphics[width=6cm, height=4cm]{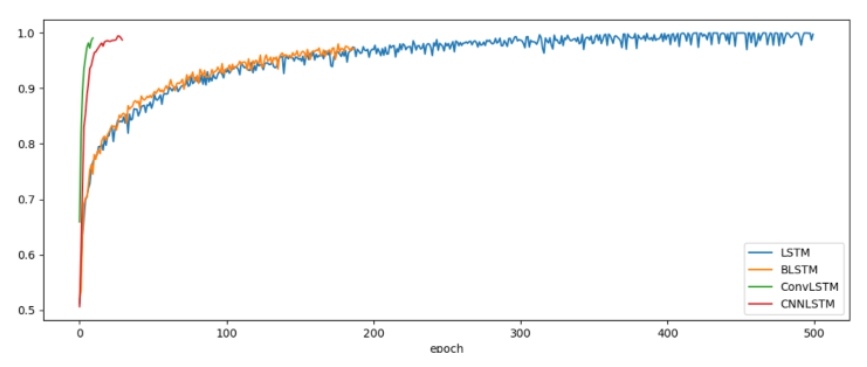}}
    \caption{Training graph of MI}
    \label{fig}
    \end{figure}
    
    \item Entropy:  

    Among the various feature extraction methods, entropy reflects the complexity of the signal. Different entropies reflect the characteristics of the signal from different views. In this paper, we propose a feature extraction method using the fusion of different entropies. The fusion can be a more complete expression of the characteristic of EEG. Entropy consisted of permutation entropy, spectral entropy, singular value decomposition entropy, approximate entropy, and sample entropy. The combination of these 5 entropies, with different Models with varying epochs(one complete pass of the training dataset after dividing them into batches) gave the best result of 99.72\% on ConvLSTM with epoch of 25, followed by 99.40\% on LSTM with epoch of 30 as mentioned in Table. 4 and Fig. 7. Also, it is clearly visible that while achieving better accuracy there was no issue of overfitting as there is not a large variation in the test and train accuracy. 
    
    \begin{figure}[htbp]
    \centerline{\includegraphics[width=6cm, height=4cm]{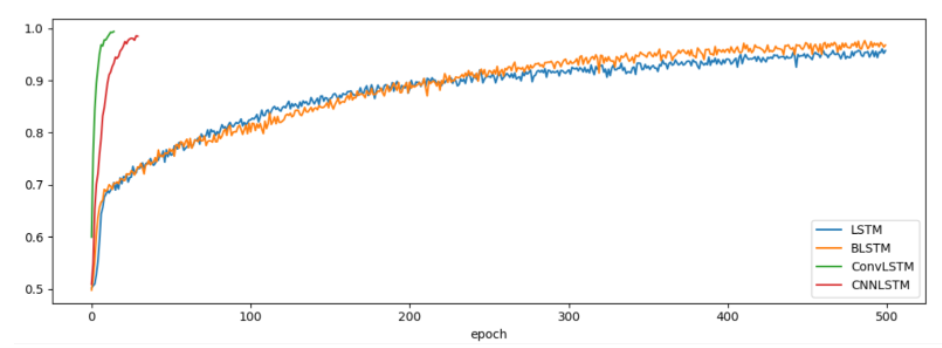}}
    \caption{Training graph of Entropy}
    \label{fig}
    \end{figure}
\end{enumerate}\par
    \begin{table}[htbp]
    \caption{Accuracy of Entropy}
    \begin{center}
    \begin{tabular}{ |p{1.5cm}|p{1.5cm}|p{2cm}|p{1cm}|  }
     \hline
     Model & Train Acc. & Test Acc. & Epoch\\
     \hline
     LSTM  & 99.40\% & 96.45 & 500 \\
     \hline
     BLSTM & 98.02\% & 95.69 & 500 \\
     \hline
     Conv-LSTM & 99.72\% & 96.66 & 25 \\
     \hline
     CNN-LSTM  & 99.34\% & 95.53\% & 30 \\
     \hline
    \end{tabular}
    \end{center}
    \end{table}
The main purpose why sample entropy and approximate was used as it helps to identify the randomness in the series of data and during the phase of calculation it can become the most influential factor. We added spectral entropy as it depicts event-related temporal change of a frequency of interest which may change when a person is doing Math. Other than these there were a couple other entropy based feature extraction like, permutation entropy which compares the neighbouring values to determine patterns and SVD. Since there are many robust measures of entropy and as a result they provided better result.

When compared to other models ConvLSTM has proven to give effective results compared to other models. When performing calculations, one can simply imagine that the states of the nieghbours which maybe unstable or the stable in a batch wise data and as a result here is why ConvLSTM stands ahead of other RNN, which simply determines the future state of a certain cell in the grid by the inputs and past states of its local neighbors. A ConvLSTM with a bigger transitional kernel should be able to catch quicker motions while one with a smaller kernel can capture slower motions if we think of the states as the hidden representations of moving objects.


    

\section{Conclusion and Future Scope}

Overall the understanding and analysis of EEG signals will help a long way in understanding the functioning of the human brain. Recently a lot of in-depth research is being done in this field. The BCS vs DCS classifier will help in classifying workload. We have extracted a lot of features and tried them on various types of classifiers. Overall the model tested on MI and Entropy showed some of the best results and entropy combined with ConvLSTM had shown an accuracy of 99.72\%. MI when tested on BLSTM showed the second-best accuracy of 99.7\%. Future work can be done on this problem using transformers for classification and extracting some complex time features. 
\newline

\vspace{12pt}

\end{document}